\documentclass[journal,article,galaxies,accept,pdftex,10pt,a4paper]{mdpi}
\usepackage{amssymb}

\firstpage{1}
\articlenumber{x}
\doinum{10.3390/------}
\pubvolume{4}
\pubyear{2016}
\copyrightyear{2016}
\externaleditor{Academic Editors: Jose L. G\'{o}mez, Alan P. Marscher and Svetlana G. Jorstad}
\history{Received: 15 July 2016; Accepted: 3 September 2016}
\pdfoutput=1

\usepackage{booktabs}
\usepackage{multirow}
\usepackage{soul}
\usepackage{microtype}
\newcommand\myurl[1]{\changeurlcolor{black}\url{#1}\changeurlcolor{blue}}
\makeatletter
\g@addto@macro{\UrlBreaks}{\UrlOrds}
\makeatother

\setitemize{parsep=6pt,itemsep=0pt,leftmargin=*,labelsep=5.5mm}
\setenumerate{parsep=6pt,itemsep=0pt,leftmargin=*,labelsep=5.5mm}
\setlist[description]{itemsep=0mm}

\Title{Applying Relativistic Reconnection to Blazar Jets}

\Author{Krzysztof Nalewajko}
\AuthorNames{Krzysztof Nalewajko}

\address[1]{Nicolaus Copernicus Astronomical Center, Polish Academy of Sciences, Bartycka 18, Warsaw 00-716 , Poland; knalew@camk.edu.pl}

\abstract{Rapid and luminous flares of non-thermal radiation observed in blazars require an~efficient mechanism of energy dissipation and particle acceleration in relativistic active galactic nuclei (AGN) jets. \mbox{Particle acceleration} in relativistic magnetic reconnection {is being actively} studied by kinetic numerical simulations. Relativistic reconnection produces hard power-law {electron energy distributions $N(\gamma) \propto \gamma^{-p}\exp(-\gamma/\gamma_{\rm max})$ with index} $p \to 1$ and exponential cut-off {Lorentz factor} $\gamma_{\rm max} \sim \sigma$ in the limit of {magnetization} $\sigma = B^2/(4\pi w) \gg 1$ {(where $w$ is the relativistic \mbox{enthalpy density})}. Reconnection in electron-proton plasma can additionally boost $\gamma_{\rm max}$ by the mass ratio $m_{\rm p}/m_{\rm e}$. Hence, in order to accelerate particles to $\gamma_{\rm max} \sim 10^6$ in the case of BL Lacs, reconnection should proceed in plasma of very high magnetization $\sigma_{\rm max} \gtrsim 10^3$. On the other hand, moderate mean jet magnetization values are required for magnetic bulk acceleration of relativistic jets, $\sigma_{\rm mean} \sim \Gamma_{\rm j} \lesssim 20$ {(where $\Gamma_{\rm j}$ is the jet bulk Lorentz factor)}. I propose that the systematic dependence of $\gamma_{\rm max}$ on blazar luminosity class---the blazar sequence---may result from a systematic trend in $\sigma_{\rm max}$ due to homogeneous loading of leptons by pair creation regulated by the energy density of high-energy external radiation fields. At the same time, relativistic AGN jets should be highly inhomogeneous due to filamentary loading of protons, which should determine the value of $\sigma_{\rm mean}$ roughly independently of the blazar class.}


\begin{document}

\section{Introduction}

Blazars are persistent extragalactic sources of non-thermal radiation extending from the radio to the gamma-ray band and characterized by stochastic variability over a wide range of time scales. Multiwavelength observations of blazars typically reveal two major broad spectral components, with the low-energy one (radio---UV/X-ray) interpreted universally as synchrotron radiation of electrons. High-resolution interferometric radio/mm observations reveal a core-jet structure with individual jet substructures propagating with apparently superluminal velocity \cite{Jor01,Lis16}. Blazars are associated with active galactic nuclei (AGN) equipped with jets (radio loud; \cite{Urr95}), with one of the jets pointing closely at the observer \cite{Bla79}, which leads to a dramatic relativistic boost of apparent luminosity \cite{Ree66}. \mbox{Blazars are} typically classified into more luminous flat-spectrum radio quasars (FSRQs) and less luminous BL Lac objects (BL Lacs). An anticorrelation between the radio luminosity and the synchrotron peak frequency is known as the blazar sequence \cite{Fos98}.

High apparent luminosities of blazars, up to $L_\gamma \sim 10^{50}\; {\rm erg\cdot s^{-1}}$ in the $\gamma$-ray band \cite{Abd11}, require efficient {in situ} dissipation of the jet power, efficient particle acceleration and efficient radiative~{mechanisms}. Modeling of the spectral energy distributions of blazars can be performed in two basic scenarios \cite{Boe13}: in the leptonic scenario, the high-energy spectral component is interpreted as inverse Compton scattering of soft radiation fields by energetic electrons; in the hadronic scenario, it is interpreted as due to various mechanisms involving relativistic protons. In luminous blazars, the requirement of high radiative efficiency favors the leptonic scenario \cite{Sik09}. However, in any case, the inferred characteristic energies of electrons producing the low-energy synchrotron component are very different in FSRQs ($\gamma_{\rm max} \sim 10^3$) and in BL Lacs ($\gamma_{\rm max} \sim 10^6$).

In this work, I attempt to address several fundamental questions about the physics of relativistic AGN jets.
\begin{enumerate} \renewcommand\labelenumi{(\theenumi)} 
\item
What is the origin of the blazar sequence? What determines $\gamma_{\rm max}$?
\item
How is matter introduced into relativistic magnetized jets? What determines $\Gamma_{\rm j}$?
\item
Can relativistic magnetic reconnection explain energy dissipation and particle acceleration in blazar jets?
\end{enumerate}

In Section \ref{recon}, I summarize the current understanding of particle acceleration in relativistic magnetic reconnection. In Section \ref{mass}, I sketch two independent mechanisms of mass loading of relativistic jets by leptons and protons. In Section \ref{seq}, I discuss why {the blazar sequence} is unlikely to be regulated by radiative cooling rates. In Section \ref{prop}, I propose a novel picture of the composition of relativistic jets and {an alternative explanation} of the blazar sequence.

\section{Particle Acceleration in Relativistic Magnetic Reconnection}
\label{recon}

Rapid progress was achieved recently in understanding particle acceleration during relativistic magnetic reconnection thanks to large-scale kinetic particle-in-cell simulations. It is now established that relativistic reconnection starting from the tearing instability of Harris-type current layers produces power-law energy distributions {with exponential cut-off $N(\gamma) \propto \gamma^{-p}\exp(-\gamma/\gamma_{\rm max})$} with index $p$ decreasing with increasing background (upstream) plasma magnetization $\sigma$ and approaching $p \to 1$ in the limit of $\sigma \gg 1$ \cite{Sir14,Guo14,Wer16}. In this limit, most energy is contained in the highest-energy particles, and hence the maximum particle energy $\gamma_{\rm max} \sim \sigma$. The high-energy cut-off in the particle energy distribution is exponential for sufficiently large simulation domains \cite{Wer16}, and no soft power-law tails that could extend $\gamma_{\rm max}$ beyond the cut-off were observed.

The acceleration mechanism was described as first-order Fermi process, in which more energetic particles achieve larger energy gains by interacting with larger plasmoids \cite{Guo14}. A theoretical explanation of the $p = 1$ electron distribution index was proposed \cite{Guo14}. Particles can gain energy at three location types: at magnetic X-points (reconnection electric field), between merging plasmoids (anti-reconnection electric field), and within accelerating plasmoids (curvature drift) \cite{Nal15}. Energy gain within contracting plasmoids \cite{Dra06} appears to be limited to non-relativistic reconnection. In Harris-layer reconnection, curvature drift has been shown to dominate global energy gain \cite{Guo14}, although the highest energy particles were found to have passed through a major X-point \cite{Sir14}. Under strong radiative cooling, the relative importance of acceleration sites changes, with rapid acceleration by strong reconnection electric fields at magnetic X-points allowing {the synchrotron photon energy to exceed the radiation reaction upper limit} \cite{Cer13}.

Relativistic reconnection in electron-ion plasma was investigated in two regimes. In the case when electrons are relativistic and ions are non-relativistic, ions gain up to 2/3 of the total dissipated energy, however their energy distributions are much softer than those of the electrons \cite{Mel14}. In the case when both electrons and ions are relativistic, they obtain similar energy distributions with hard power-law spectra \cite{Guo16}, hence this case is qualitatively similar to reconnection in relativistic pair plasma. In any case involving relativistic electrons, their maximum Lorentz factor is given by $\gamma_{\rm max} \sim \sigma_{\rm e} \sim (n_{\rm p}m_{\rm p}/n_{\rm e}m_{\rm e})\sigma_{\rm p} \lesssim 10^3\sigma_{\rm p}$ (Werner et~al., in preparation).

An alternative numerical setup was investigated, starting from smooth magnetostatic equilibria dubbed ``ABC'' fields \cite{Nal16,Yua16,Lyu16}. Such an initial condition allows one to study the formation and dynamical evolution of the current layer and simultaneous particle acceleration and radiation. In ABC reconnection, energy gain by the reconnection electric field appears to be more important.

\section{Mass Loading in Relativistic Jets}
\label{mass}

Energy can be extracted from rotating black holes by purely electromagnetic jets \cite{Bla77}, hence, there is no fundamental principle for the presence of matter at the base of jets. Without introducing matter from outside, there is no fundamental upper limit on the value of magnetization $\sigma$. Here I consider mass loading by pairs and protons as independent mechanisms.

Mass loading by pairs can be due to photon-photon pair creation. This requires the presence of roughly isotropic soft gamma-ray radiation or interaction of roughly isotropic X-ray radiation with relativistically boosted X-ray radiation of the jet \cite{Sik00}. If this mechanism can operate efficiently in the first place, it can be expected to operate uniformly across the jet volume, providing a lower limit on plasma density, and hence an upper limit on magnetization $\sigma$. Since this mechanism is very sensitive to the energy density of external high-energy radiation fields, the upper limit on $\sigma$ can be expected to depend systematically on blazar type. In the case of FSRQs, we expect the pair creation to be more efficient, and the maximum magnetization value to be lower, as compared with the BL Lacs.

On the other hand, mass loading by protons can be due to penetration of the jet by gas clouds or even stars \cite{Bar10} or due to various plasma instabilities, e.g., the magnetic Kelvin-Helmholtz instability or magnetic Rayleigh-Taylor instability \cite{McK12}. These processes can be expected to be highly non-uniform, leading to high density contrasts with most of the protons concentrated in compact clouds/filaments/comae/tails. We may also expect a significant radial stratification of the jets, with the inner cores essentially free of protons, unless the entire jet is subject to a global current driven~instability. This will likely result in high contrasts of magnetization $\sigma$. The proton loading~mechanisms can be expected to be independent of the radiative environment of the jet. Whether they should depend on the jet power or the mechanical jet environment, is an interesting open question.

\section{On the Origin of Blazar Sequence}
\label{seq}

The blazar sequence is an observational correlation between the frequency of the spectral peak of the synchrotron component $\nu_{\rm syn}$ and the synchrotron luminosity measured in the radio band~\cite{Fos98}. \mbox{It should} be noted that the observational evidence for the reality of blazar sequence is being questioned as due to selection {effects} \cite{Gio12}. Nevertheless, for individual blazars it is quite striking that despite dramatic variability in all bands their broad-band SEDs remain stable. Given similar values of the co-moving magnetic field strength $B'$ and the jet bulk Lorentz factor $\Gamma_{\rm j}$, the synchrotron peak frequency is determined by the characteristic energy of radiating electrons $\nu_{\rm syn} \propto \gamma_{\rm peak}^2$. By modeling the spectral energy distributions of blazars belonging to different classes, a theoretical anticorrelation between $\gamma_{\rm peak}$ and the combined co-moving energy density of magnetic fields and external radiation fields \linebreak $u_{\rm cool}' = u_{\rm B}'+u_{\rm ext}'$ was demonstrated \cite{Ghi98,Cel08}: $\gamma_{\rm peak} \sim 10^3 (u_{\rm cool,0}')^{-0.5}$, where $u_{\rm cool,0}'~=~u_{\rm cool}'/(1\;{\rm erg \cdot cm^{-3}})$. It was then proposed that the blazar sequence arises from acceleration of electrons being controlled by their radiative cooling rate $t_{\rm acc}' \sim t_{\rm cool}'$.

Acceleration of electrons under radiative cooling is governed by the following equation:
\begin{equation}
\frac{{\rm d}\gamma}{{\rm d}t'} = \frac{eE'}{m_{\rm e}c} - \frac{4\sigma_{\rm T}u_{\rm cool}'\gamma^2}{3m_{\rm e}c}\,,
\end{equation}
where $E'$ is the effective electric field parallel to the instantaneous particle velocity vector. In the particular case of relativistic magnetic reconnection, but also in any other acceleration mechanism, we can relate $E'$ to the local magnetic field strength as $E' = \beta B'$. In the case of reconnection, $\beta \equiv \beta_{\rm rec}$ is the reconnection rate, and in other cases it can be interpreted as a characteristic relative velocity of interacting magnetized flows. The equilibrium electron energy is found simply from ${\rm d}\gamma / {\rm d}t' = 0$ to be
\begin{equation}
\gamma_{\rm eq} = \sqrt\frac{3e\beta B'}{4\sigma_{\rm T}u_{\rm cool}'} \simeq 2\times 10^7\sqrt\frac{\beta B_0'}{u_{\rm cool,0}'}\,,
\end{equation}
where $B_0' = B'/(1\;{\rm G})$. This relation predicts that $\gamma_{\rm eq} \propto u_{\rm cool}'^{-0.5}$, but only when radiative cooling is dominated by Comptonization of the external radiation and when $B_0' \sim {\rm const}$. However, in order to match the empirical relation between $\gamma_{\rm peak}$ and $u_{\rm cool}'$, with $B_0' \sim 1$ in parsec-scale jets \cite{Pus12}, one would require $\beta \sim 10^{-9}$. Even considering the maximum energy of the particle distributions obtained from SED modeling $\gamma_{\rm max} \sim 10\gamma_{\rm peak}$, one would need $\beta \sim 10^{-7}$. In other words, \emph{for the maximum electron energies to be determined by radiative losses, the acceleration mechanism should be extremely slow.} \mbox{Relativistic magnetic} reconnection, with $\beta_{\rm rec} \sim 0.1$ \cite{Liu15}, should easily accelerate electrons to the energies of order $\gamma_{\rm max} \sim 10^7$. In such {a case}, the SEDs of blazars would be more similar to that of the Crab Nebula with the synchrotron component extending up to $\sim$100\;${\rm MeV}$ \cite{Jag96}. An analogous problem in the context of particle acceleration by relativistic shocks was identified by \cite{Ino96}.

\section{Proposition}
\label{prop}

Here I propose a qualitatively new picture of blazars and relativistic jets. At given distance scale, relativistic jets are composed of roughly homogeneous magnetic energy density, roughly homogeneous pair density, and highly inhomogeneous proton density. The magnetic field and the proton components are roughly independent of the blazar class, and the pair component scales with the high-energy external radiation density $u_{\rm ext,he}$. The pair density determines the local maxima of the jet magnetization $\sigma_{\rm max}$ decreasing with increasing $u_{\rm ext,he}$, and the proton density determines the mean jet magnetization $\sigma_{\rm mean}$ roughly independent of $u_{\rm ext,he}$. The maximum jet magnetization determines the maximum energy of electrons $\gamma_{\rm max}$, potentially explaining the blazar sequence. The mean jet magnetization determines the jet bulk Lorentz factor $\Gamma_{\rm j}$, which depends only very weakly on the blazar class.

This model provides a suitable background for relativistic reconnection as the main dissipation mechanism in blazars. Without the existence of very highly magnetized regions, with $\sigma_{\rm max} \sim 10^3$, acceleration of electrons at reconnection sites cannot produce the energy distributions required for TeV BL Lacs.
{The existence of very highly magnetized regions in relativistic jets is not inconsistent with the dominant emitting regions being roughly in equipartition, since in relativistic reconnection the emitting regions are in general expected to be in sharp contrast to the main acceleration regions \cite{Sir15}. In any case, blazar emission produced in highly magnetized relativistic jets would be expected to be characterized by very high synchrotron luminosity and low Compton dominance \cite{Jan15}.}

\section{Conclusions}
\label{con}

Relativistic reconnection is a promising dissipation mechanism in relativistic jets. Rapid progress in understanding particle acceleration during relativistic reconnection has been achieved recently by means of kinetic numerical simulations. We {have learned} that relativistic reconnection can accelerate particles very efficiently, producing hard power-law distributions with index $p = 1$ and exponential cut-off at $\gamma_{\rm max} \sim \sigma_{\rm e}$. Applying this mechanism to blazars requires very high jet magnetization values, $\sigma_{\rm max} \sim 10^3$ in the case of TeV BL Lacs. I suggest that such high magnetizations may be present locally in relativistic jets due to highly inhomogeneous mass loading of protons by plasma instabilities. Fast~reconnection rates $\beta_{\rm rec} \sim 0.1$ can easily compete with radiative cooling rates, allowing in principle {acceleration of} particles to much higher energies than required by SED modeling, especially in the case of FSRQs. Therefore, I argue that the blazar sequence can hardly be regulated by radiative \mbox{cooling rates}. Instead, I suggest that the blazar sequence arises due to homogeneous loading of leptons by pair creation regulated by external radiation fields. Several aspects of the proposed picture of blazar physics require detailed theoretical or numerical verification.

\vspace{6pt}

\acknowledgments{I thank Marek Sikora and Mitch Begelman for stimulating discussions. This work was supported by the Polish National Science Centre grant 2015/18/E/ST9/00580.}

\conflictofinterests{The author declares no conflict of interest.}

\pagebreak

\bibliographystyle{mdpi}

\begin{thebibliography}{999}

\bibitem[{Jorstad} {et~al.}(2001)]{Jor01}
{Jorstad}, S.G.; {Marscher}, A.P.; {Mattox}, J.R.; {Wehrle}, A.E.; {Bloom}, S.D.; {Yurchenko}, A.V.
Multiepoch very long baseline array observations of EGRET-detected quasars and BL lacertae objects: Superluminal motion of gamma-ray bright blazars.
{\em Astrophys. J. 	Suppl. } {\bf 2001}, {\em 134}, 181--240.

\bibitem[Lister et~al.(2016)]{Lis16}
Lister, M.L.; Aller, M.F.; Aller, H.D.;  Homan, D.C.;  Kellermann, K.I.;  Kovalev, Y.Y.;  Pushkarev, A.B.;  \mbox{Richards, J.L.}; Ros, E.;  Savolainen, T.
MOJAVE: XIII. Parsec-scale AGN jet kinematics analysis based on \mbox{19 years} of VLBA observations at 15 GHz.
{\em Astron. J.} {\bf 2016}, {\em 152}, 12.

\bibitem[Urry \& Padovani(1995)]{Urr95}
Urry, C.M.; Padovani, P.
Unified schemes for radio-loud active galactic nuclei.
{\em Publ. Astron. Soc. Pac.} {\bf 1995}, {\em 107}, 803--845.

\bibitem[{Blandford} \& {K\"onigl}(1979)]{Bla79}
{Blandford}, R.D.; {K\"onigl}, A.
Relativistic jets as compact radio sources.
{\em Astrophys. J.} {\bf 1979}, {\em 232}, 34--48.

\bibitem[Rees(1966)]{Ree66}
Rees, M.J.
Appearance of relativistically expanding radio sources.
{\em Nature} {\bf 1966}, {\em 211}, 468--470.

\bibitem[{Fossati} {et~al.}(1998)]{Fos98}
{Fossati}, G.; {Maraschi}, L.; {Celotti}, A.; {Comastri}, A.; {Ghisellini}, G.
A unifying view of the spectral energy distributions of blazars.
{\em Mon. Not. R.  Astron. Soc.} {\bf 1998}, {\em 299}, 433--448.

\bibitem[{Abdo} {et~al.}(2011)]{Abd11}
{Abdo}, A.A.; {Ackermann}, M.; {Ajello}, M.;  Allafort, A.;  Baldini, L.;  Ballet, J.;  Barbiellini, G.; Bastieri, D.;   Bellazzini, R.;  Berenji, B.; {et~al.}
Fermi gamma-ray space telescope observations of the gamma-ray outburst from 3C454.3 in November 2010.
{\em Astrophys. J.} {\bf 2011}, {\em 733}, L26.

\bibitem[B{\"o}ttcher et~al.(2013)]{Boe13}
B{\"o}ttcher, M.; Reimer, A.; Sweeney, K.; Prakash, A.
Leptonic and hadronic modeling of fermi-detected blazars.
{\em Astrophys. J.} {\bf 2013}, {\em 768}, 54.

\bibitem[{Sikora} {et~al.}(2009)]{Sik09}
{Sikora}, M.; {Stawarz}, \L.; {Moderski}, R.; {Nalewajko}, K.; {Madejski}, G.M.
Constraining emission models of luminous blazar sources.
{\em Astrophys. J.} {\bf 2009}, {\em 704}, 38--50.

\bibitem[Guo et~al.(2014)]{Guo14}
Guo, F.; Li, H.; Daughton, W.; Liu, Y.-H.
Formation of hard power laws in the energetic particle spectra resulting from relativistic magnetic reconnection.
{\em  Phys. Rev. Lett.} {\bf 2014}, {\em 113}, 155005.

\bibitem[Sironi \& Spitkovsky(2014)]{Sir14}
Sironi, L.; Spitkovsky, A.
Relativistic reconnection: An efficient source of non-thermal particles.
{\em Astrophys. J.} {\bf 2014}, {\em 783}, L21.

\bibitem[Werner et~al.(2016)]{Wer16}
Werner, G.R.; Uzdensky, D.A.; Cerutti, B.; Nalewajko, K.; Begelman, M.C.
The extent of power-law energy spectra in collisionless relativistic magnetic reconnection in pair plasmas.
{\em Astrophys. J.} {\bf 2016}, {\em 816}, L8.

\bibitem[Nalewajko et~al.(2015)]{Nal15}
Nalewajko, K.; Uzdensky, D.A.; Cerutti, B.; Werner, G.R.; Begelman, M.C.
On the distribution of particle acceleration sites in plasmoid-dominated relativistic magnetic reconnection.
{\em  Astrophys. J.} {\bf 2015}, {\em 815}, 101.

\bibitem[Drake et~al.(2006)]{Dra06}
Drake, J.F.; Swisdak, M.; Che, H.; Shay, M.A.
Electron acceleration from contracting magnetic islands \mbox{during reconnection}.
{\em Nature} {\bf 2006}, {\em 443}, 553--556.

\bibitem[Cerutti et~al.(2013)]{Cer13}
Cerutti, B.; Werner, G.R.; Uzdensky, D.A.; Begelman, M.C.
Simulations of particle acceleration beyond the classical synchrotron burnoff limit in magnetic reconnection: An explanation of the crab flares.
{\em Astrophys. J.} {\bf 2013}, {\em 770}, 147.

\bibitem[Melzani et~al.(2014)]{Mel14}
Melzani, M.; Walder, R.; Folini, D.; Winisdoerffer, C.; Favre, J.M.
Relativistic magnetic reconnection in collisionless ion-electron plasmas explored with particle-in-cell simulations.
{\em Astron.  Astrophys.} {\bf 2014}, {\em 570}, A111.

\bibitem[Guo et~al.(2016)]{Guo16}
Guo, F.; Li, X.; Li, H.;  Daughton, W.;  Zhang, B.;  Lloyd-Ronning, N.;  Liu, Y.-H.;  Zhang, H.;  Deng, W.
\mbox{Efficient production} of high-energy nonthermal particles during magnetic reconnection in a magnetically dominated ion-electron plasma.
{\em Astrophys. J.} {\bf 2016}, {\em 818}, L9.

\bibitem[Lyutikov et~al.(2016)]{Lyu16}
Lyutikov, M.; Sironi, L.; Komissarov, S.; Porth, O.
Particle acceleration in explosive relativistic reconnection events and crab nebula gamma-ray flares. {\bf 2016}, arXiv:1603.05731.

\bibitem[Nalewajko et~al.(2016)]{Nal16}
Nalewajko, K.; Zrake, J.; Yuan, Y.; East, W.E.; Blandford, R.D.
Kinetic simulations of the lowest-order unstable mode of relativistic magnetostatic equilibria.
{\em Astrophys. J.} {\bf 2016}, {\em 826}, 115.

\bibitem[Yuan et~al.(2016)]{Yua16}
Yuan, Y.; Nalewajko, K.; Zrake, J.; East, W.E.; Blandford, R.D.
Kinetic study of radiation-reaction-limited particle acceleration during the relaxation of unstable force-free equilibria.
{\bf 2016}, arXiv:1604.03179.

\bibitem[Blandford \& Znajek(1977)]{Bla77}
Blandford, R.D.; Znajek, R.L.
Electromagnetic extraction of energy from Kerr black holes.
{\em Mon. Not. R.  Astron. Soc.} {\bf 1977}, {\em 179}, 433--456.

\bibitem[Sikora \& Madejski(2000)]{Sik00}
Sikora, M.; Madejski, G.
On pair content and variability of subparsec jets in quasars.
{\em Astrophys. J.} {\bf 2000}, {\em 534}, 109--113.

\bibitem[{Barkov} {et~al.}(2010)]{Bar10}
{Barkov}, M.V.; {Aharonian}, F.A.;  {Bosch-Ramon}, V.
Gamma-ray flares from red giant/jet interactions in active galactic nuclei.
{\em Astrophys. J.} {\bf 2010}, {\em 724}, 1517--1523.

\bibitem[McKinney et~al.(2012)]{McK12}
McKinney, J.C.; Tchekhovskoy, A.; Blandford, R.D.
General relativistic magnetohydrodynamic simulations of magnetically choked accretion flows around black holes.
{\em Mon. Not. R.  Astron. Soc.} {\bf 2012}, {\em 423}, 3083--3117.

\bibitem[Giommi et~al.(2012)]{Gio12}
Giommi, P.; Padovani, P.; Polenta, G.; Turriziani, S.;  D'Elia, V.;   Piranomonte, S.
A simplified view of blazars: Clearing the fog around long-standing selection effects.
{\em Mon. Not. R.  Astron. Soc.} {\bf 2012}, {\em 420}, 2899--2911.

\bibitem[{Celotti} \& {Ghisellini}(2008)]{Cel08}
{Celotti}, A.; {Ghisellini}, G.
The power of blazar jets.
{\em Mon. Not. R.  Astron. Soc.} {\bf 2008}, {\em 385}, 283--300.

\bibitem[Ghisellini et~al.(1998)]{Ghi98}
Ghisellini, G.; Celotti, A.; Fossati, G.; Maraschi, L.; Comastri, A.
A theoretical unifying scheme for gamma-ray bright blazars.
{\em Mon. Not. R.  Astron. Soc.} {\bf 1998}, {\em 301}, 451--468.

\bibitem[Pushkarev et~al.(2012)]{Pus12}
Pushkarev, A.B.; Hovatta, T.; Kovalev, Y.Y.;  Lister, M.L.;  Lobanov, A.P.;  Savolainen, T.;  Zensus, J.A. MOJAVE: Monitoring of jets in active galactic nuclei with VLBA experiments. IX. Nuclear opacity. {\em Astron.  Astrophys.} {\bf 2012}, {\em 545}, A113.

\bibitem[Liu et~al.(2015)]{Liu15}
Liu, Y.-H.; Guo, F.; Daughton, W.; Li, H.; Hesse, M. Scaling of magnetic reconnection in relativistic collisionless pair plasmas. {\em Phys. Rev. Lett.} {\bf 2015}, {\em 114}, 095002.

\bibitem[{de~Jager} {et~al.}(1996)]{Jag96}
{De~Jager}, O.C.; {Harding}, A.K.; {Michelson}, P.F.;  Nel, H.I.; Nolan, P.L.; Sreekumar, P.; Thompson, D.J.
Gamma-ray observations of the crab nebula: A study of the synchro-compton spectrum.
{\em Astrophys. J.} {\bf 1996}, {\em 457}, 253--266.

\bibitem[Inoue \& Takahara(1996)]{Ino96}
Inoue, S.; Takahara, F.
Electron acceleration and gamma-ray emission from blazars.
{\em Astrophys. J.} {\bf 1996}, {\em 463}, 555--564.

\bibitem[Sironi et~al.(2015)]{Sir15}
Sironi, L.; Petropoulou, M.; Giannios, D.
Relativistic jets shine through shocks or magnetic reconnection?
{\em Mon. Not. R.  Astron. Soc.} {\bf 2015}, {\em 450}, 183.

\bibitem[Janiak et~al.(2015)]{Jan15}
Janiak, M.; Sikora, M.; Moderski, R.
Magnetization of jets in luminous blazars.
{\em Mon. Not. R.  Astron. Soc.} {\bf 2015}, {\em 449}, 431--439

\end{thebibliography}
\renewcommand\bibname{References}

\end{document}